\documentclass[prb,amssymb,twocolumn,showpacs]{revtex4}

\usepackage{amssymb}
\usepackage{graphicx}
\usepackage{bm}

\begin{document}

\title{ Large Amplitude Harmonic Driving of Highly Coherent Flux Qubits}

\author{Alejandro Ferr\'on}
\author{Daniel Dom\'{\i}nguez}
\author{Mar\'{\i}a Jos\'{e} S\'{a}nchez}

\affiliation{Centro At{\'{o}}mico Bariloche and Instituto Balseiro,
8400 San Carlos de Bariloche,
R\'{\i}o Negro, Argentina.}

\begin{abstract}
The device for the Josephson flux qubit (DJFQ)  can be considered as a solid state
artificial atom with multiple energy levels. When a large amplitude harmonic excitation is applied to the system,
transitions at the energy levels avoided crossings produce visible changes in  the qubit population over many driven periods that are accompanied by a rich
pattern of interference phenomena. We present a Floquet treatment of the periodically
 time-dependent Schr\"odinger equation of the strongly driven   qubit beyond the standard two
 levels approach.  For low amplitudes,  the average probability of a given sign of the persistent
 current qubit exhibits, as a function of the static flux detuning and the driving amplitude,
Landau-Zener-St\"uckelberg (LZS) interference patterns that evolve  into complex diamond-like patterns for large amplitudes.
In the case of highly coherent flux qubits  we show that the higher order diamonds
can not be simply described relying on LZS transitions in each avoided crossing
considered separately.
In addition we propose a new spectroscopic method based on starting the
 system in the first excited state instead of in the ground state, which can give further information
 on the energy level spectrum and dynamics in the case of highly coherent flux qubits.
We compare our numerical results with recent experiments that
perform amplitude spectroscopy to probe the energy spectrum of the artificial atom.
\end{abstract}


\maketitle

\section{Introduction}

In recent years, several types of superconducting qubits have been
experimentally studied.\cite{nakamura,qbit_mooij,vion,martinis,revqubits} These
systems consist on mesoscopic Josephson devices  and constitute
promising candidates to be used for the design of qubits for
quantum
computation.\cite{nakamura,qbit_mooij,vion,martinis,revqubits,chiorescu,fqubit_recent,noise}
Indeed, a large effort is devoted to succeed in the coherent
manipulation of their quantum states in a controlable way. The
progress made along this line allows to have nowadays Josephson
circuits with small dissipation and large decoherence
times.\cite{vion,martinis,chiorescu,fqubit_recent,noise}

In this work we will focus on  the device for the Josephson flux
qubit (DJFQ), which consists of a SQUID loop with three Josephson
junctions operated at or near a magnetic flux of half
quantum.\cite{qbit_mooij,chiorescu,fqubit_recent,noise}. When
cooling down to millikelvin temperatures this device exhibits
quantized levels whose energies can be tuned by a control
parameter such as an exernal magnetic field. This   artificial
atom-like behavior has motivated several  studies  based on the
analysis of  the level spectrum  and its dynamics  beyond the
simplified two-level approach.
 As an example it has been shown  that,
after the inclusion of  higher energy levels  the  DJFQ
exhibits  quantum signatures  of classical chaos;\cite{mingo,pozzo}
and a recent study \cite{ferron} focused on the calculation of the
intrinsic leakage ( i.e. transitions from the allowed qubit states
to higher excited levels of the system) has shown that for very
strong  resonant harmonic pulses the two-level approximation
breaks down.

What is more important, several recent experiments driving the
flux qubit with a combination of a dc and  large amplitude
harmonic excitations in the magnetic flux have studied the
energy level structure through Landau-Zener-St\"uckelberg
transitions.\cite{izmalkov,oliver,izmalkov2,valenzuela} Mach-Zender
interferometry\cite{oliver,izmalkov2} and amplitude spectroscopy
\cite{valenzuela} have been  the subject of these recent
experimental studies of the flux qubit as an artificial atom.
In particular, the amplitude spectroscopy experiment of
Ref.\onlinecite{valenzuela} has revealed
the higher energy level spectrum when increasing the microwave
amplitude. In this case, the
average population of one state of the DJFQ as a function of the
dc flux (flux detuning) and microwave amplitude  exhibits
diamond-like interference patterns, which display a rich
structure of multi-photon resonances.\cite{oliver}
From   these interference patterns it is in principle possible to reconstruct a
large fraction of the energy spectrum and methods based on two
dimensional Fourier transform have been recently proposed  to this
end. \cite{rudner}
In the experiment, the observed spectroscopic
``diamonds'' arise due to  combined contributions of
Landau-Zener-St\"uckelberg transitions, which provide the
interference fingerprint of  different energy level avoided
crossings, together with intra-well fast relaxation and short
coherence times, which provide contrast in the observed pattern.

Recent theoretical efforts have been put forward to reconstruct
the experimentally observed interference patterns by solving the
dynamics of the model under strong driving. Most of the reported
approaches reduce the model for the DJFQ to  a simplified version
which only considers the dynamics of the two levels involved
in each avoided crossing. In this case
 the  well known Landau Zener- St\"uckelberg theory has been applied,
considering only the accumulated phase of the two levels during a period of the driving.\cite{oliver,shevchenko} The beginning of the first
spectroscopic diamond, that corresponds to the first avoided
crossing, is  accurately  reproduced within this basic
model.\cite{oliver,shevchenko} Additionally, extensions  that
incorporate  several levels, based on rate equations  along all
coupled levels have been  recently proposed.\cite{wen} In this
later case more than one diamond can be obtained, but the approach
neglects the phase accumulated in the evolution of several levels
and can only be applied when decoherence and relaxation effects
are important. Since this case is somewhat near the experiment of
Ref.\onlinecite{valenzuela}, a qualitative description of the
observed diamond patterns can be obtained. However, the devices
for the flux qubit  can have larger decoherence
times\cite{noise,nota} than in the case of
Ref.\onlinecite{valenzuela}. In this case, the effect of fast
intra-well relaxation that provided contrast in the diamond
patterns, and the effect  of short decoherence times that made
possible to consider only the accumulated phases of two-levels at
the avoided crossings, will be much weaker. Then the question
arises on how the interference patterns of strongly driven DJFQ
can be analyzed in the highly coherent case, and how much
information on the energy spectrum can be extracted in this
situation.

The purpose of this work is to solve the dynamics of the DJFQ
under strong driving rf pulses considering the full hamiltonian of
the system. We perform a first-principles calculation taking
as input only two parameters from the experimental device:
the ratio of the Josephson and charging energies, $E_J/E_C$, and
the asymmetry factor $\alpha$ of the Josephson energy of one of the
junctions with respect to the others.
Furthermore, our approach focus on the behavior of highly coherent
DJFQ when driven within time scales smaller than the dephasing
time. Therefore the interaction with the environment is neglected
and we solve the time dependent Schr\"odinger equation considering
the DJFQ as a closed system. As we will show, even when relaxation and
dephasing are neglected, our results reproduce several of the qualitative
features of the experiment of Ref.\onlinecite{valenzuela}.

We will employ the Floquet formalism \cite{floquet} which has been extensively
applied to study time dependent periodic evolutions in systems
ranging from two level systems, including simplified models of
flux qubits, \cite{son} to more realistic molecular and nanoscaled
systems \cite{variosfloq}.
The Floquet method allows to transform the periodically time
dependent Schr\"{o}dinger equation into an equivalent infinitely
dimensional eigenvalue problem for a time independent Floquet
matrix. In general several truncations schemes are employed  in
order to tackle the analytical solution and reduce the infinite
Floquet matrix to an effective finite dimensional matrix.\cite{ho}

The paper is organized as follows. In Sec. \ref{model}  we introduce the model
Hamiltonian and equations for the Josephson flux qubit.
In Sec. \ref{sn1} we present numerical results for the amplitude
spectroscopy for the Josephson flux qubit by direct numerical calculation and
using the Floquet formulation  for the time dependent Schr\"{o}dinger equation
in the case of an harmonic drive. In this section we compare our numerical
results with recent experimental realizations. In Sec. \ref{sn2} we propose
a new amplitude spectroscopy method by changing the initial conditions.
Numerical calculation using the Floquet formulation are presented. Finally,
Sec. \ref{sc} contains a summary and a discussion of the most relevant
points of our findings.

\section{Model for the Device for the Josephson Flux Qubit}
\label{model}

\begin{figure}[th]
\begin{center}
\includegraphics[width=0.8\linewidth]{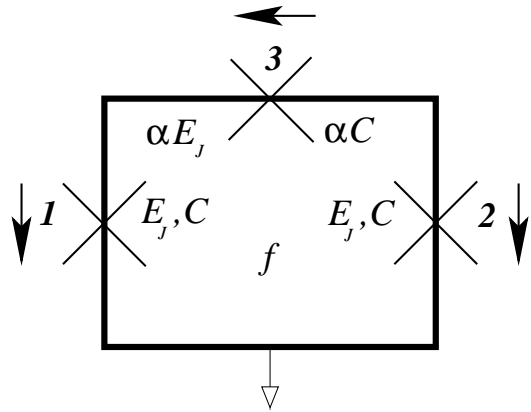}
\end{center}
\caption{Circuit for the DJFQ as described in the text.
Josepshon junctions $1$ and $2$ have Josepshon energy $E_J$ and capacitance $C$, and junction $3$
has Josepshon energy and capacitance $\alpha$ times smaller.
The arrows indicate the sign
convention for defining the gauge invariant phase differences. The circuit encloses
a magnetic flux $\Phi = f \Phi_0$.}
\label{djfq_fig}
\end{figure}

The DJFQ  consists on a superconducting ring with  three Josephson
junctions\cite{qbit_mooij} enclosing a magnetic flux $\Phi=
f\Phi_0$, with $\Phi_0=h/2e$, see Fig.\ref{djfq_fig}.

The junctions have gauge invariant phase differences defined as
$\varphi_1$, $\varphi_2$ and $\varphi_3$, respectively, with the
sign convention corresponding to the directions indicated by the
arrows in Fig.\ref{djfq_fig}. Typically the circuit inductance can
be neglected  and the phase difference of the third junction is:
$\varphi_3=-\varphi _1 +\varphi _2 -2\pi f$. Therefore the system
can be described with  two dynamical variables:
$\varphi_1,\varphi_2$. The circuits that are used for the DJFQ
have two of the junctions with the same coupling energy,
$E_{J,1}=E_{J,2}=E_J$, and capacitance, $C_1=C_2=C$, while the
third junction has smaller coupling $E_{J,3}=\alpha E_J$ and
capacitance $C_3=\alpha C$, with $0.5<\alpha<1$. In terms of the
two-dimensional coordinate $\vec{\varphi}= (\varphi_1,\varphi_2)$,
the hamiltonian of the DJFQ is: \cite{qbit_mooij}

\begin{equation}
\label{hamil}
{\cal H}=\frac{-\eta^2}{2} \nabla_\varphi^T{\rm{\bf m}}^{-1}\nabla_\varphi
+V(\vec {\bf \varphi})\; ,\label{ham_clas}
\end{equation}
where we have normalized  ${\cal H}$ by the Josephson coupling energy $E_J$,
and   $\eta^2=8E_C/E_J$, with  $E_C=e^2/2C$.
The kinetic term of the hamiltonian corresponds to the electrostatic energy
of the system, where the momentum operator is
 $\hat{\vec{P}}= -i\hbar \nabla_\varphi$, with
$\nabla_\varphi=(\frac{\partial}{\partial\varphi_1},\frac{\partial}{\partial\varphi_2})$,
and the ``mass'' tensor is given by the matrix ${\rm{\bf m}}$,
$$
{\rm {\bf m}}=\left(
{{\begin{array}{cc}
 {1+\alpha  }  & {-\alpha }  \\
 {-\alpha }  & {1+\alpha  }  \\
\end{array} }} \right).$$

The potential term of the hamiltonian corresponds to the Josephson energy
of the junctions, and is given by
\begin{equation}
\label{eq:pot}
V(\vec {\bf \varphi})=
2+\alpha -\cos \varphi_1-\cos \varphi_2
- \alpha \cos (2\pi f+\varphi _1 -\varphi _2 ) \; .
\end{equation}

Typical  flux qubit experiments  have values of $\alpha$ in the
range $0.6-0.9$ and $\eta$ in the range $0.1-0.6$.
\cite{chiorescu,fqubit_recent,noise,valenzuela}

In quantum computation implementations
\cite{qbit_mooij,chiorescu,fqubit_recent} the Josephson flux qubit
is operated at (static) magnetic fields near the half-flux quantum,  $f=
1/2+\delta f$, with $\delta f \ll 1$. For $\alpha \ge
1/2$, the potential of Eq.(\ref{eq:pot}) has the shape of a
double-well with two minima (within the domain
$-\pi<\varphi_1<\pi$, $-\pi<\varphi_2<\pi$). After a change of
variables one can define the transverse phase
$\varphi_t=(\varphi_1+\varphi_2)/2$ and the longitudinal phase
$\varphi_l=(\varphi_1-\varphi_2)/2$, obtaining
\begin{equation}
\label{eq:pot2} V(\vec {\bf \varphi})= 2+\alpha -
2\cos\varphi_t\cos \varphi_l - \alpha \cos (2\pi f+2\varphi _l) \;
.
\end{equation}
The two minima are along the longitudinal direction $\varphi_l$,
at $(\varphi_l,\varphi_t)=(\pm\varphi^*,0)$ separated by a
maximum at $(\varphi_l,\varphi_t)=(0,0)$. Each minima corresponds
to macroscopic persistent currents of opposite sign. Experimental
measurements are sensitive to the sign of the persistent
current,\cite{chiorescu} and therefore they detect the probability
of being on one side or the other of the double-well potential.
For $\delta f> 0$ ($\delta f < 0$) a ground state $|-\rangle$
($|+\rangle$) with negative (positive) persistent current is
favored, with energy $\epsilon_\pm\propto\pm\delta f$. At $\delta
f=0$ the two minima have the same energy, and the two lowest
energy eigenstates ($|\Psi_0\rangle$ and $|\Psi_1\rangle$) are
symmetric and antisymmetric superpositions of the two states
($|-\rangle$ and $|+\rangle$)) corresponding to the macroscopic
persistent currents. To describe the dynamics of the device as a
quantum bit, a two-level truncation of the Hilbert space is
performed. \cite{qbit_mooij} In the subspace expanded by
$|\Psi_0\rangle$ and $|\Psi_1\rangle$, the hamiltonian of
Eq.~(\ref{hamil}) is reduced to
\begin{equation}\label{htls}
 {\cal H}_{TLS}=-\frac{\epsilon}{2} {\hat\sigma}_z - \frac{\Delta}{2} {\hat\sigma}_x
\;,
\end{equation}
where ${\cal H}_{TLS}$ is written in the basis defined by
$|+\rangle=(|\Psi_0\rangle+|\Psi_1\rangle)/\sqrt{2}$ and
$|-\rangle=(|\Psi_0\rangle-|\Psi_1\rangle)/\sqrt{2}$. Here
$\Delta=E_1-E_0$ is the two-level splitting at $\delta f=0$, and
$\epsilon= 4\pi\alpha E_J S_{01} \delta f$ (for $\delta f\ll1$),
with $S_{01}=-\langle\Psi_0|\sin(2\varphi _l) |\Psi_1\rangle
=-\langle + |\sin(2\varphi _l) |+\rangle$. (For typical values of
$\alpha$ and $\eta$, one has  $S_{01}\sim 0.8-0.9$). Most
experiments control the system varying the magnetic field detuning
$\delta f$. The magnitude of the gap $\Delta$ depends
exponentially on $\alpha$ and $\eta$. Recently it has been shown
experimentally that it is possible to manipulate the value of
$\Delta$ by controlling $\alpha$, replacing the third junction by
an additional SQUID loop.\cite{mooij09,shimazu}

Landau-Zener-St\"uckelberg (LZS) interferometry is performed by
applying an harmonic field on top of the static field  such that $f\rightarrow f(t)$ with
\begin{equation}\label{tdf}
 f(t)=f_0+f_{p}\sin{(\omega t)} \;.
\end{equation}
Hence Eq.(\ref{htls}) acquires an  explicit dependence on time
through $\epsilon \rightarrow \epsilon (t)= \epsilon_{0} + A \sin
(\omega t)$ with $\epsilon_{0}= 4\pi\alpha E_J S_{01} \delta f$,
$\delta f=f_0-1/2$ and $A\equiv  4 \pi\alpha E_{J}  S_{01} f_p$.
The initial state corresponds to  prepare the  system   in the
ground state $|0,f_0\rangle$ for  the static field  $f_0$.

For values of $|\delta f|\ll1$ and small driving amplitudes
$f_p\ll1$, the DJFQ  is adequately described as  a two level
system (TLS), whose time evolution  under an harmonic drive does
not have, in general, an exact solution. Thus the dynamics is
usually approximated\cite{shevchenko,ashhab} by free evolutions of
the basis states mediated by  non adiabatic Landau-Zener (LZ)
transitions \cite{zener}, with probability  $P_{LZ}= \exp{(- 2\pi
\delta)}$ with  ${\delta}= \Delta^{2}/ f_{p} \omega$. In the last
case, explicit expressions for the occupation probability have
been obtained in the fast (slow) driving
regime,\cite{shevchenko,ashhab}  $\delta \ll (\gg) 1$.

Here we will consider the case of nearly fast driving,
which corresponds to the series of experiments on
amplitude spectroscopy performed in  Ref.\onlinecite{valenzuela}.
For $f_0\lesssim 1/2$, the system is started in the
ground state $|0,f_0\rangle\approx |+\rangle$, which has
a positive persistent current. In this type of experiments,
one asks for the probability of switching to a state of negative
persistent current: $P_{|+\rangle\rightarrow|-\rangle}=P_{-}(t)$
during the time the harmonic pulse is applied.
For the TLS in the fast driving regime, the occupation
probability $P_{-}$ is approximately given as:
\cite{shevchenko,ashhab}
\begin{eqnarray}\label{tfd}
 P_{-}^{TFD}(t) & = & \sum_{n} \frac{\Gamma_{n}^{2}}{2 {\Omega}_{n}^{2}}
\left((1 - \cos ({\Omega}_{n}  t)\right) \; , \label{p2ls}\\
\Gamma_{n} &=&   \Delta \; J_{n} (A/ \omega) \; , \nonumber \\
\Omega_{n} &= & \sqrt{(n \omega - {\epsilon}_{0})^{2} +
\Gamma_{n}^{2} } \;,\nonumber
\end{eqnarray}
being $J_{n} (z)$ the order $n$ Bessel function of the first kind.
The resonance condition $\epsilon_0 = n \omega$ is attained  when the
total phase accumulated over a single period of the driving,
$\Theta= 2 \pi \epsilon_0 / \omega$,  satisfies  $ \Theta =  2 \pi n $
 for a given integer $n$.\cite{oliver,shevchenko}  Under resonance, the occupation probability
 $P_-^{TFD}(t) \rightarrow 1/2 \left( 1- \cos (\Omega_{n} t)\right)$ with
$\Omega_{n}=\Delta J_{n} (A/\omega)$. Notice that
$\Omega_{n}$ depends  on the driving amplitude $f_p$  through $A$.

Besides the time dependence,  the average occupation probability  is the key
quantity for  the spectroscopic analysis performed in  recent experiments.  \cite{valenzuela,oliver}
In the case of a TLS in the fast driving regime,  the average occupation probability  obtained
from Eq.(\ref{tfd}) is a sum of Lorentzian-shape $n-$ photon resonances\cite{oliver}

\begin{equation}
 \overline{P_{-}^{TFD}}=\frac{1}{2}\sum_n\frac{\Gamma_{n}^{2}}
{(n \omega - {\epsilon}_{0})^{2} + \Gamma_{n}^{2}}  \; . \label{l2ls}
\end{equation}

\noindent Thus as  $\epsilon_{0}$ (or $\delta f$) is changed,
different n-resonances  are explored. In addition the Bessel
function entering in $\Gamma_{n}$ gives a quasiperiodic character
to the patterns of resonances as the amplitude $f_{p}$ is varied
keeping the frequency $\omega$ fixed.

The analysis of the positions of the resonances  as a function
of $f_{p}$ and  $\delta f$ was the route followed in Refs.\onlinecite{valenzuela,oliver}
 in an effort to obtain the parameters characterizing the different avoided crossings
of the flux qubit.

\section{Amplitude Spectroscopy for Coherent Systems}
\label{sn1}
\subsection{Direct numerical calculation}
\label{sn1a}

In this section we will focus on the study of  the quantum dynamics
of the DJFQ  driven by the time dependent flux $f(t)$ given in
Eq.(\ref{tdf}) for  $f_0$ near $1/2$ and varying the amplitude $f_p$ of the harmonic
drive.

In the absence of driving, i.e for $f_{p}=0$, the
eigenvectors   $\Psi_{n}(\vec {\bf \varphi})$  and eigenenergies $E_n$
are obtained by solving,
 \begin{equation}
\label{eq:Schro}
 \left[ -\frac{\eta^2}{2}\nabla_\varphi^T{\rm{\bf m}}^{-1}\nabla_\varphi
 +V(\vec {\bf \varphi})\right] \Psi_n(\vec {\bf \varphi}) =
E_n \Psi_n(\vec {\bf \varphi})\;.
 \end{equation}

\begin{figure}[th]
\includegraphics[width=15pc]{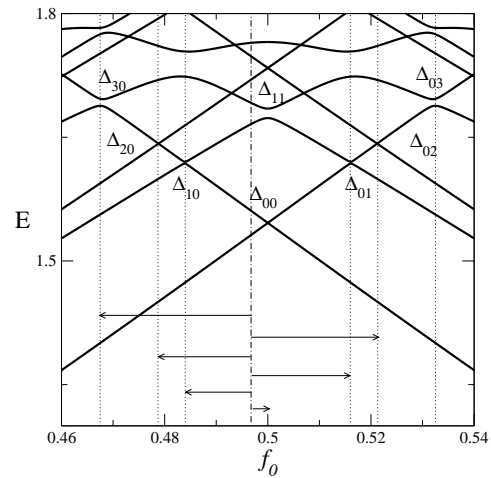}
\caption{Lowest seven  energy levels  of the DJFQ as a function of
flux $f_{0}$ for  $\eta=0.25$ and $\alpha=0.80$. Arrows indicate
the position  of the avoided level crossings $\Delta_{ij}$
measured from $f_{0}=0.497$ (indicated by the vertical dashed
line). Calculations were done using $M=256$ (see the text for
details). Energy is measured in units of $E_J$ and flux in units
of $\Phi_0$.} \label{fig2}
\end{figure}

In Fig.(\ref{fig2}) we plot the seven lower energy levels  as a
function of flux detuning $f_{0}$, obtained  by numerical
diagonalization of  Eq.~(\ref{eq:Schro}) using a discretization
grid of $\Delta\varphi=2\pi/M$  and $2\pi$-periodic boundary
conditions on $\vec {\bf\varphi}=(\varphi_1,\varphi_2)$. In this
case we set $\eta= 0.25$  and $\alpha=0.8$, close to the
experimental values  employed in flux qubits experiments.
\cite{qbit_mooij,valenzuela} The  energy spectrum is rather
sensitive to the values of $\eta$ and $\alpha$, and in particular
for the selected values, the energy landscape is quite involved,
presenting many avoided crossings  $\Delta_{ij}$ in the range
$0.45<f_{0}<0.55$. The slope of the energy levels $dE_n/df_{0}$ is
proportional to the average current in the loop. Therefore an
eigenstate with positive or negative slope, corresponds to a wave
function mostly weighted in one side or the other of the
double-well. A gap $\Delta_{ij}$ opens at the avoided crossings of
energy levels of opposite slope. We label the gaps $\Delta_{ij}$
as the avoided crossing of the $i$-th level of positive slope with
the $j$-th level of negative slope, see Fig.(\ref{fig2}). (This
convention is different from  the one used in
Ref.\onlinecite{valenzuela} where a distinction among longitudinal
and transverse modes is made in the labeling of the gaps.)

It is evident from the energy level diagram  of Fig.\ref{fig2} that the description of the
time evolution of the  DJFQ in terms of a TLS is  valid  only for a
very  small range of amplitudes $f_{p}$.

In the presence of a finite driving amplitude $f_p$, our   first approach to the problem is to
solve numerically the  time dependent
Schr\"{o}dinger equation (we have normalized time by $t_J =\hbar/E_J$)
\begin{equation}
i\frac{\partial\Psi(\vec {\bf \varphi},t)}{\partial t}={\cal H}\Psi(\vec {\bf \varphi},t) \; .
\label{eqst-b}
\end{equation}
 We integrate
numerically Eq.~(\ref{eqst-b}) with  a second order split-operator
algorithm, \cite{feit} using a discretization grid of
$\Delta\varphi=2\pi/M$ and $\Delta t= 0.1 t_J$. We use
$2\pi$-periodic boundary conditions on $\vec {\bf
\varphi}=(\varphi_1,\varphi_2)$. The system is started in the
ground state $|0,f_0\rangle$ for a given static field $f_0$, obtained
from the numerical solution of Eq.~(\ref{eq:Schro}).
Experimentally, what is measured is the probability of being in a
given state of positive, $P_+$, or negative, $P_-$, persistent
current, which can be obtained as:
\begin{equation}\label{pp}
P_+(t)\equiv 1-P_-(t)=\int_{\pi>\varphi_l>0}
|\Psi(\varphi_1,\varphi_2,t)|^2 d\varphi_1d\varphi_2 \; ,
\end{equation}
\begin{figure}[th]
\includegraphics[width=20pc]{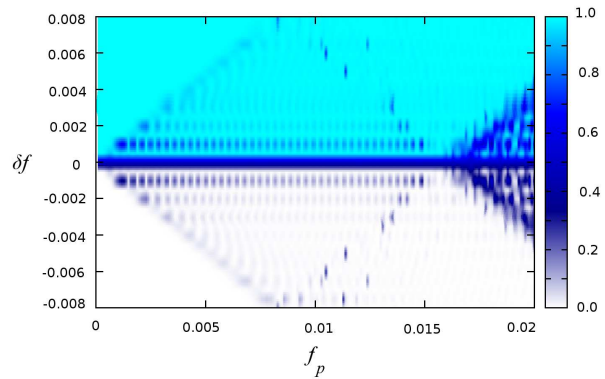}
\caption{(Color online) Large amplitude spectroscopic diamonds
obtained for the Josephson flux qubit with $\eta=0.25$ and
$\alpha=0.8$. The intensity of $\overline {P}_{-}$ is plotted as a
function of flux detuning $\delta f$ and rf amplitude $f_p$. The
system is driven at a frequency $\omega=0.001$ (in units of
$E_J/\hbar$). Numerical calculations were done by direct numerical
integration using $M=128$ and $\Delta t=0.1 t_J$.
Data points correspond to a grid of $\Delta f_0 = 1 \times10^{-3} $ and $\Delta f_p =1
\times10^{-4}$.
} \label{fig2c}
\end{figure}

\noindent where the integration  is on one side of the double-well
potential defined  by $\varphi_l>0$ (i.e., $-\pi\leq\varphi_1\leq
\pi$ and $\varphi_2 \leq \varphi_1$). The quantity measured
experimentally is the long time occupation probability, which
is equivalent to the  time average probability in the stationary state.
In Fig.\ref{fig2c} we
plot the time averaged probability $\overline{P}_-=1-\overline{P}_+$ as a function of
the static flux $f_0$ and the amplitude $f_p$ of the harmonic
excitation. The average is performed over several periods
of the harmonic drive (typically $\sim 20 - 100$ periods, until
convergence of the average).
The plot is obtained by calculating points with a grid
of $\Delta f_0 = 1 \times10^{-3} $ and $\Delta f_p =1
\times10^{-4}$. A pattern of ``spectroscopic diamonds'' is
observed, similar to the one obtained in the experiments, which
can be related to the energy level spectrum of Fig.\ref{fig2} as
follows. At a fixed flux detuning $\delta f = f_0 -1/2$, the first
diamond, D1, starts when the $\Delta_{00}$ avoided crossing is
reached, at $f_p=f_{1s}=f_{00}-f_0=-\delta f$, with $f_{00}=1/2$
the location of $\Delta_{00}$. The first diamond ends when the
$\Delta_{10}$ crossing is reached at $f_p=f_{1e}=f_0-f_{10}$, with
$f_{10}$ the location of $\Delta_{10}$. Then the second diamond,
D2, starts when the $\Delta_{01}$ avoided crossing is reached, at
$f_p=f_{2s}=f_{01}-f_0$, with $f_{01}$ the location of
$\Delta_{01}$, etc. The spectroscopic diamonds of Fig.\ref{fig2c}
have much less contrast than in the experiments of
Ref.\onlinecite{valenzuela} in which the contrast is due to fast
intra-well relaxation. This induces population inversion, reducing
the population in the sectors between the diamonds (i.e. for
example between D1 and D2). In DJFQs with less relaxation effects,
the picture should be closer to the one shown in Fig.\ref{fig2c}.
Within the first diamond a regular pattern of resonances can be
qualitatively observed in Fig.\ref{fig2c}. However to accurately
describe all the resonances of D1, as well as the complex
structure of D2, a finer grid sampling $\Delta f_0,\Delta f_p$ is
needed. Furthermore, simulations at high $f_p$, in the region of
the second diamond and above (where the  dynamics has more weight
in higher energy levels), need a better discretization of the
Schr\"{o}dinger equation and averaging of the population for
larger times.
Therefore a finer description of the
structure of diamonds needs large time consuming simulations of
the full time dependent Schr\"{o}dinger equation. Instead, in the
following we will employ an approach based on the Floquet
formalism, more adequate for time periodic hamiltonians.

\subsection{The Floquet Formulation} \label{ff}
\label{sn1b}

For a finite driving $f(t)$, we  write
${\cal H}=\ {\cal H}_{0}+ \delta V(\vec{\varphi},t)$ with ${\cal H}_{0}$ corresponding
to Eq.~(\ref{hamil}) with $f=f_0$ (i.e. the time independent part of the hamiltonian) and
\begin{eqnarray}
\delta V(\vec{\varphi},t)&=&\alpha E_J\sin[2\pi f_{ac}(t)]\sin(2\pi f_0 +\varphi_1-\varphi_2)+\nonumber\\
&&\!\!\!2\alpha E_J\sin^2[\pi f_{ac}(t)]\cos(2\pi f_0
+\varphi_1-\varphi_2)\;, \label{vt1}
\end{eqnarray}

\noindent where $f_{ac}(t)=f_{p}\sin{(\omega t)}$. The potential
defined in Eq.(\ref{vt1}) is periodic in time  with period
$T=2\pi/\omega$. Thus, according to the Floquet theorem,
\cite{floquet} the time dependent Schr\"{o}dinger equation (\ref{eqst-b})
has a solution that can be written as

\begin{equation}
\Psi_\alpha(t)=e^{-i\varepsilon_\alpha t} \Phi_\alpha(t) \;,
\label{flo1}
\end{equation}

\noindent where $\Phi_\alpha(t)=\Phi_\alpha(t+T)$ and
$\varepsilon_\alpha$ is known as the quasienergy or Floquet
eigenvalue.

Substituting expression (\ref{flo1}) into Eq.(\ref{eqst-b})
we obtain an eigenvalue equation for the quasienergies:

\begin{equation}
\hat{H}_F(t) \Phi_\alpha(t) =\varepsilon_\alpha  \Phi_\alpha(t) \; ,
\label{flo2}
\end{equation}
\noindent where the Floquet Hamiltonian is defined as

\begin{equation}
\hat{H}_F(t)={\cal H}(t)-i\frac{\partial}{\partial t} \; .
\label{flo3}
\end{equation}

Since the function $\Phi_\alpha(t)$ is periodic in time it can  be
expanded  in a Fourier series. We introduce  the standard Floquet
nomenclature\cite{variosfloq} and define $| n,k\rangle = |
n\rangle \otimes | k\rangle$, where $n$ is an index that labels
the eigenstates of $H_{0}$ and $k$ is a Fourier index. Then

\begin{equation}
\langle n|\Phi_\alpha(t)\rangle=\sum_{k=-\infty}^\infty \langle n,k|\phi_\alpha
\rangle e^{-ik\omega t}
\label{flo6}
\end{equation}

\noindent where $\langle n,k|\phi_\alpha\rangle$ is a Fourier amplitude.
From Eq.(\ref{flo2}-\ref{flo6}) it is straightforward to  write

\begin{equation}
\varepsilon_\alpha \langle n,q|\phi_\alpha\rangle = \sum_{m}\sum_{k}
\langle n,q|\hat{H}_F|m,k\rangle\langle m,k|\phi_\alpha\rangle
\label{flo7}
\end{equation}

\noindent where $\hat{H}_F$ is the Floquet Hamiltonian previously defined
whose matrix elements are given by

\begin{eqnarray}
\langle n,q|\hat{H}_F|m,k\rangle=(E_n+q\omega)\delta_{m,n}\delta_{k,q}+
\nonumber \\
\frac{\omega}{2\pi} \int_{0}^{2\pi/\omega}
\langle n\left|\delta V(\vec{\varphi},t)\right|m\rangle
e^{-i(q-k)\omega t} dt \; .
\label{flo8}
\end{eqnarray}

\noindent Then the time dependent problem is reduced to solve the eigenvalue equation
Eq.(\ref{flo7}).

We  need to calculate the matrix elements of the Floquet
Hamiltonian defined in Eq.(\ref{flo8}). As it was mentioned
before, the first term in this equation is obtained by numerical
diagonalization of Eq.~(\ref{eq:Schro}), using a discretization
grid of $\Delta\varphi=2\pi/M$ and $2\pi$-periodic boundary
conditions on $\vec {\bf\varphi}=(\varphi_1,\varphi_2)$.  The
second term in
 Eq. (\ref{flo8}) is  written  as

\begin{equation}
V_{nm}^{l}=\frac{\omega}{2\pi}\int_{0}^{2\pi/\omega}
\langle n\left|\delta V(\vec{\varphi},t)\right|m\rangle
e^{-il\omega t}  dt \;,
\label{flo9}
\end{equation}

\noindent where $\delta V(t)$ was defined in Eq.(\ref{vt1}) and $l=q-k$ is an integer.
The integration is straightforward and we obtain

\begin{eqnarray}
V_{nm}^{l}= \alpha E_J \times \left \{ \begin{array}{ll}
                       {\cal C}_{nm}(\delta_{l0}-J_{l}(2\pi f_p)) & \mbox{for {\it l} even} \\
                        i{\cal S}_{nm}J_{l}(2\pi f_p) & \mbox{for {\it l} odd} \; ,
            \end{array}
\right.
\label{flo12}
\end{eqnarray}

\noindent where ${\cal S}_{nm}=\langle n|\sin{(2\pi f_{0}+\varphi_1-
\varphi_{2})}|m\rangle$, ${\cal C}_{nm}=\langle n|
\cos{(2\pi f_{0}+\varphi_1-\varphi_{2})}|m\rangle$ and
 $J_{l}(x)$ is the Bessel function of first kind of order $l$.

Then  we have all the ingredients to construct the Floquet matrix

\begin{equation}
\langle n,q|\hat{H}_F|m,k\rangle=(E_n+q\omega)\delta_{m,n}\delta_{k,q}+
V_{nm}^{q-k} \;,
\label{flo13}
\end{equation}

\noindent where $q$ and $k$ range over all integers form $-\infty$
to $\infty$. In order to solve the problem numerically we must
truncate the Floquet matrix, Eq.(\ref{flo13}). The truncated
matrix  is  of dimension $N_d =(2K+1)N_l$ where $K$ is
defined by the maximum value of the Fourier index and $N_l$  by
the number of levels considered in  the diagonalization of Eq.
(\ref{eq:Schro}).

Floquet eigenstates and quasienergies contain all the information
 to construct the  large amplitude spectroscopic diamonds.
Following the experiments, we take as initial state the ground
state of ${\cal H}_{0}$ for a given value of flux detuning $f_{0}$
that here for simplicity, we denote  $|0\rangle$. The initial
state is prepared at a time $t_0$, and then at a
time $t$ the evolved solution  $|\Psi(t,t_0)\rangle$ can be expanded
in the basis of eigenstates of ${\cal H}_{0}$ as
\begin{equation} \label{wf}
\Psi(\varphi_1,\varphi_2,t,t_0)=\sum_{n}c_n(t,t_0)
\chi_n(\varphi_1,\varphi_2)
\end{equation}
\noindent where
$\chi_n(\varphi_1,\varphi_2)=\langle\varphi_1,\varphi_2 |n\rangle$
is the  wave function representation of eigenket $|n\rangle$ in
terms of the variables $(\varphi_1,\varphi_2)$.

Using the Fourier amplitudes of the Floquet eigenstates
$ \langle n,k|\phi_\alpha\rangle$,
and their corresponding eigenenergies $\varepsilon_\alpha$,
one obtains the coefficients\cite{variosfloq}
\begin{equation}
c_{n}(t,t_0)=\sum_k\sum_\beta \langle
n,k|\phi_\beta\rangle\langle \phi_\beta|0,0\rangle
e^{-i\varepsilon_\beta (t-t_0)} e^{i\omega k t}
\label{cnt}
\end{equation}
which are the probablity amplitudes that the system initialy in the ground state at time
$t_0$ evolves to a state $|n\rangle$ by time $t$ according to the time-periodic
Hamiltonian. This equation can be interpreted as the amplitude probability
that the system initially in the Floquet state $|0,0\rangle$ at time $t_0$
evolve to the Floquet state $|n,k\rangle$ by time $t$ according to the time
independent Floquet Hamiltonian, sumed over $k$ with weighting factors
$\exp{(i\omega k t)}$.

We can now calculate the time dependence of the probability $P_+$ (or  $P_-=1-P_+$)
of a measurement of a positive (negative)  state of persistent current,
replacing
Eqs.(\ref{wf}) and (\ref{cnt}) into Eq.(\ref{pp}), and obtaining,

\begin{equation}\label{p+}
P_+(t,t_0)=\sum_{n}\sum_{m}\lambda_{nm}(t,t_0)
p_{nm} \; ,
\end{equation}

\noindent where the coefficients

\begin{equation}
\label{pnm}
p_{nm}= \int_{{\cal W}}\chi_n(\varphi_1,\varphi_2)
\chi_m^\ast(\varphi_1,\varphi_2) d\varphi_1d\varphi_2 \; ,
\end{equation}
\noindent are evaluated by numerical integration using the eigenstates
of Eq.(\ref{eq:Schro}) and ${\cal W}$ is the triangular sector of the two
dimensional space defined  by $\varphi_l>0$ (i.e.,
$-\pi\leq\varphi_1\leq \pi$ and $\varphi_2 \leq \varphi_1$).
The time dependent coefficients $\lambda_{nm}(t,t_0)=c_n(t,t_0)c_m^\ast(t,t_0)$
are calculated as

\begin{eqnarray}\label{ltt0}
&\lambda_{nm}(t,t_0)=\sum_{k,l}\sum_{\beta,\gamma} \langle
n,k|\phi_\beta\rangle\langle \phi_\beta|0,0\rangle\nonumber \\
&\langle 0,0|\phi_\gamma
\rangle\langle \phi_\gamma|m,l\rangle
e^{-i(\varepsilon_\gamma-\varepsilon_\beta) (t-t_0)} e^{i\omega(k - l)t} \; .
\end{eqnarray}

\noindent
We can calculate now the time average of $P_+(t)$.
To this end, we average $\lambda_{nm}(t,t_0)$  over many periods
considering that  for long times
$\overline{e^{i[\omega(k-l)+\varepsilon_\gamma-\varepsilon_\beta]t}}=
\delta_{\omega(k-l),\varepsilon_\beta-\varepsilon_\gamma} $.
Using the periodic properties of the quasienergies and
eigenvectors of the Floquet Hamiltonian,\cite{variosfloq} we get
\begin{eqnarray}\label{ltt1}
&\overline{\lambda}_{nm}(t_0)=\sum_{k,l}\sum_{\beta} \langle
n,k|\phi_\beta\rangle\langle \phi_\beta|0,0\rangle\nonumber \\
&\langle 0,k-l|\phi_\beta
\rangle\langle \phi_\beta|m,k\rangle
e^{i\omega(k -
 l)t_0}
\end{eqnarray}
\noindent
In experiments the initial time, or equivalently the initial phase
of the field seen by the system in repeated realizations of the
measurement, is not well defined. Then, the quantity of
interest is the transition probability averaged over initial times,\cite{variosfloq}
in which case $ \overline{e^{i\omega(k-l)t_0}}=
\delta_{k,l}$. Finally, we obtain

\begin{equation}
\overline{P}_+=\sum_{n,m} p_{nm} \Lambda_{nm}\; ,
\label{p+a}
\end{equation}

\noindent where

\[
\Lambda_{nm}=\sum_{\beta}\sum_{k} \langle
n,k|\phi_\beta\rangle\langle \phi_\beta|0,0\rangle \langle
0,0|\phi_\beta\rangle\langle \phi_\beta|m,k\rangle  \;.
\]

\noindent Thus  once equiped with the  Floquet  quasienergies and eigenstates
one can either  compute  the time dependent occupation probability Eq.(\ref{p+}) or the
time average probability Eq.(\ref{p+a}), which is indeed  an exact average.

To summarize, we follow this procedure:
(i) We take the parameters $\eta$ and $\alpha$ from
the experimental device. (Here we use  $\eta=0.25$ and $\alpha=0.8$.)
(ii) For each value of the magnetic flux $f_0$  we solve numerically the eigenvalue
equation (\ref{eq:Schro}) obtaining the eigenstates $\chi_n(\varphi_1,\varphi_2)$
and eigenvalues $E_n$, with a discretization $\Delta\varphi=2\pi/M$ ($M=256-1024$).
(iii) We evaluate the Floquet matrix elements of Eq.(\ref{flo12}) and the
coefficients $p_{nm}$ of Eq.(\ref{pnm}).
(iv) We solve numerically the Floquet eigenvalue equation, Eqs.(\ref{flo7}) and (\ref{flo13})
obtaining the Floquet eigenvalues
$\varepsilon_\alpha$ and the components  $ \langle n,q|\phi_\alpha\rangle$ of the
corresponding eigenvectors.
A truncation of the matrix is performed: we consider $N_l$ energy levels
(results for different $N_l$ will be shown), and we consider $2K+1$ Fourier components in $-K < k < K$.
Large $K$ is chosen until convergence of the quantities of interests.($K \sim  150-400$)
(v) The time dependent occupation probability Eq.(\ref{p+}) for $t_0=0$,  or the
time average probability Eq.(\ref{p+a})  are then evaluated, with the sums over energy levels
between $0$ and $N_l-1$, the sums over $k$ components between $-K$ and $K$, and the
sums over Floquet states between $1$ and $N_d=N_l(2K+1)$.

\subsubsection{Time Dependent Occupation Probabilities}
\label{sn1c}

\begin{figure}[th]
\begin{center}
\includegraphics[width=20pc]{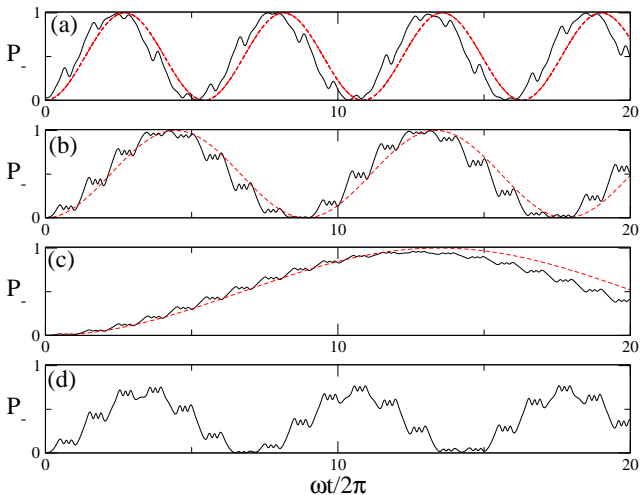}
\end{center}
\caption{(Color online) $P_{-} $ as a function of time $\tau= \omega t /2 \pi$ for  $\eta=0.25$ and $\alpha=0.8$.
 (a) $\epsilon_{0}= \omega$, for  $\omega=0.001$,
$\delta f= -0.11 \times10^{-3}$, and $f_{p}=0.24 \times10^{-3}$ 
(black solid line).
The fast driving TLS Eq.(\ref{p2ls}) for $n=1$ with
$\Omega_{1}= \Delta_{00} J_{1}(4\pi S_{01} f_p/\omega)=1.87 \times10^{-4}$
is plotted for comparison (red dashed line).
(b)$\epsilon_{0}= 3 \omega$
for $\omega=0.001$, $\delta f= -0.33 \times10^{-3} $ and  $f_{p}=0.35 \times10^{-3}$
(black solid line). The fast driving TLS Eq.(\ref{p2ls}) for $n=3$ with
$\Omega_3= \Delta_{00} J_{3}(4\pi S_{01} f_p/\omega)=1.14 \times10^{-4}$
is plotted for comparison (red dashed line).
(c)  $\epsilon_{0}= 3 \omega$ for  a higher frequency $\omega=0.003$,
$\delta f= -0.99 \times10^{-3} $ and  $f_{p}=1.05 \times10^{-3}$
(black solid line). Notice that the oscillations on time scales
 $\sim \pi/ \omega$ are smaller for  $\omega=0.003$, as described  in the text.
The fast driving TLS Eq.(\ref{p2ls}) for $n=3$  is plotted for comparison
(red dashed line).
 (d) Out of ($n=3$) resonance,  for $\omega=0.001$, $\delta f= -0.32 \times10^{-3}$.
 $f_{p}=0.35 \times10^{-3}$  (black solid line).
Notice that the oscillations on time scales  $\sim \pi/\omega$
persist. Numerical calculations were done with $N_l=6$ levels,
$K=180$ and $M=1024$. Calulations with $N_l=2$ levels overlap
almost exactly with the $N_l=6$ calculations in this case.} \label{fig3}
\end{figure}

We start analyzing the explicit time dependence of the probability
$P_{-} (\tau)$ after a driving  of duration $\tau$ is applied.
To this end, we evaluate numerically Eq.(\ref{p+}) for $t_0=0$.
Different  initial states  shall correspond to prepare the  system
in the ground state $|n=0;f_0\rangle$ for different  $f_0 \lesssim
0.5$.

For values of $\delta f=f_0-1/2$ and (rather small) driving
amplitudes $f_p$, such that    $ \Delta_{00}$  is the  only
relevant avoided crossing (see Fig.\ref{fig2}), the DJFQ  can be
described as  a TLS. In Fig.\ref{fig3}  we plot $P_{-} (\tau)= 1 -
P_{+} (\tau)$ as a function of time  $ \tau =  \omega t / 2 \pi $
in units of the pulse period, for small $f_p$. The numerical
calculations were performed  with the Floquet formalism, employing
in Eq.(\ref{p+}) the lowest six energy levels ($N_l=6$) and
$K=150-250$. For this case, we find that calculations
with $N_l=2$ levels overlap almost exactly with the $N_l=6$ calculations,
and can not be distinguished in the plot, since
the TLS approximation is correct for small $f_p$, as expeceted.
In panel (a) we consider the case with frequency
$\omega=0.001$ and   with $\delta f= -0.11 \times10^{-3}$ such that
it corresponds to the  fast driving TLS $n=1$ resonance, $\epsilon_{0}=
\omega$. For the selected amplitude, $f_{p}=0.25 \times10^{-3}$,
the behavior of  $P_{-} (\tau)$ is on the  global scale rather
well described by  the fast driving approximation for a TLS given by Eq.(\ref{tfd}).
The numerically obtained frequency is very close to
 $\Omega_{1} \sim \Gamma_{1}=1.87 \times10^{-4} $, in agreement with the  on -resonance  relation
written in  Eq.(\ref{p2ls}).
However, $P_{-} (\tau)$ exhibits sudden jumps mediated by  additional oscillations with $n$ local
maxima on time scales $\sim \pi/\omega$.
These oscillations  reflect the quantum mechanical  interference between consecutive passages through
the avoided crossing $\Delta_{00}$.
This behavior is not captured by the fast  driving TLS expression
Eq.(\ref{p2ls}) plotted for comparison by the red dashed line.  In the present case
$\delta=\Delta^{2}_{00}/\omega \ f_{p} \simeq0.5$  and then,
the interference effects are important.
In panel (b) we consider  for the same frequency as in (a), the case of $\delta f= -0.33\times10^{-3}$
to select the fast driving TLS   $n=3$ resonance
($\epsilon_{0}= 3 \omega$).
For a small amplitude $f_{p}=0.35 \times10^{-3}$,
the  qualitative description given for panel (a) holds. For the driving parameters
used in panel (b) we have an adiabaticity parameter $\delta\simeq 0.25$.
In this case the oscillations on time scales $\sim \pi/\omega$, exhibit $n=3$ local maxima.
In panel (c) we also consider the case of the $n=3$ resonance  but at a higher
frequency, $\omega=0.003$, for which $\delta\simeq 0.03$. In this case the fast
driving approximation is more adequate, the local oscillations on time scales $\sim \pi/\omega$
are washed out, improving the agreement with  the  fast driving TLS expression Eq.(\ref{p2ls}).
This can be understood taking into account that
the  (adiabatic) Landau-Zener transition probability   \cite{zener} at a single
avoided crossing diminishes as the  frequency $\omega$ increases.
In Fig.\ref{fig3} (d) $P_{-} (\tau)$ is plotted  for  $\delta f= -0.32 \times10^{-3}$
and $\omega=0.001$,  a value which is out but
close to the $n=3$ resonance. We employ  the same amplitude $f_{p}= 0.35 \times10^{-3}$
as in Fig.\ref{fig3} (b).
Notice that in this case is  $ \max[ {P_{-} (\tau)}] <1$, as expected in the off-resonance
situation, but
the short time scale  oscillations on time scales $\sim \pi/\omega$ persist.

\begin{figure}[th]
\begin{center}
\includegraphics[width=20pc]{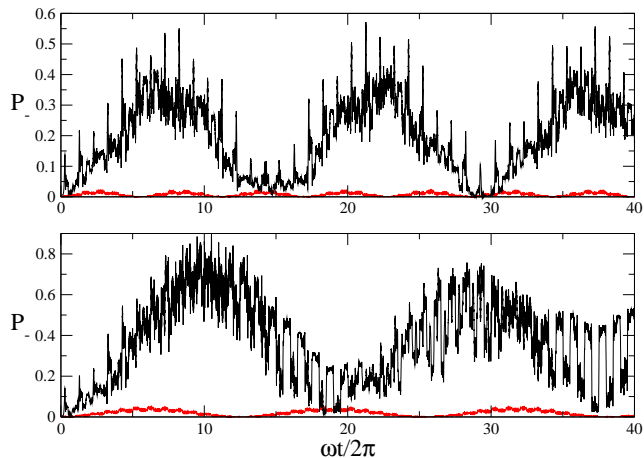}
\end{center}
\caption{(color online) $P_{-} $ as a function of time $\tau= \omega t /2 \pi$ for  $\eta=0.25$
and $\alpha=0.8$.  The flux qubit is driven at a strong  amplitude
with $f_p=16 \times10^{-3}$.
(a) resonance condition for the TLS in the fast
driving regime:  $\epsilon_{0}= 3 \omega$ for $\omega=0.001$,
 $\delta f= -0.33 \times10^{-3} $.
  (b) $\epsilon_{0}\neq n \omega$,  for $\omega=0.001$,
$\delta f= -0.32 \times10^{-3}$. Numerical calculations were done
with $N_l=6$ levels, $K=300$ and $M=1024$.
Calculations with $N_l=2$ levels are plotted for comparison (red
line).} \label{fig4}
\end{figure}

In Fig.\ref{fig4} we show that for a larger amplitude $f_{p}= 16
\times10^{-3}$ the TLS  approach breaks down. For this amplitude,
the system is driven close to the avoided crossing $\Delta_{01}$
(see Fig.\ref{fig2}). In panel (a) we show the case where
$\epsilon_{0}= 3 \omega$ is satisfied while in panel (b) we show the case
out of the  $\epsilon_{0}= 3 \omega$ condition. As expected, the
magnitude of $\max [{ P_{-} (\tau)}]$ in this case is completely
unrelated with the resonance conditions observed at smaller $f_p$.
We also compare the results obtained considering up to $N_l=6$
levels when evaluating Eq.(\ref{p+})  with the case with only
$N_l=2$ levels. As it is evident in the plots,  for this large
amplitude  more than two levels are needed to describe the
behavior of $P_{-} (\tau)$, since most of the population is at the
higher energy eigenstates.

\subsubsection{Average Occupation Probabilities}
\label{sn1d}

\begin{figure}[th]
\begin{center}
\includegraphics[width=20pc]{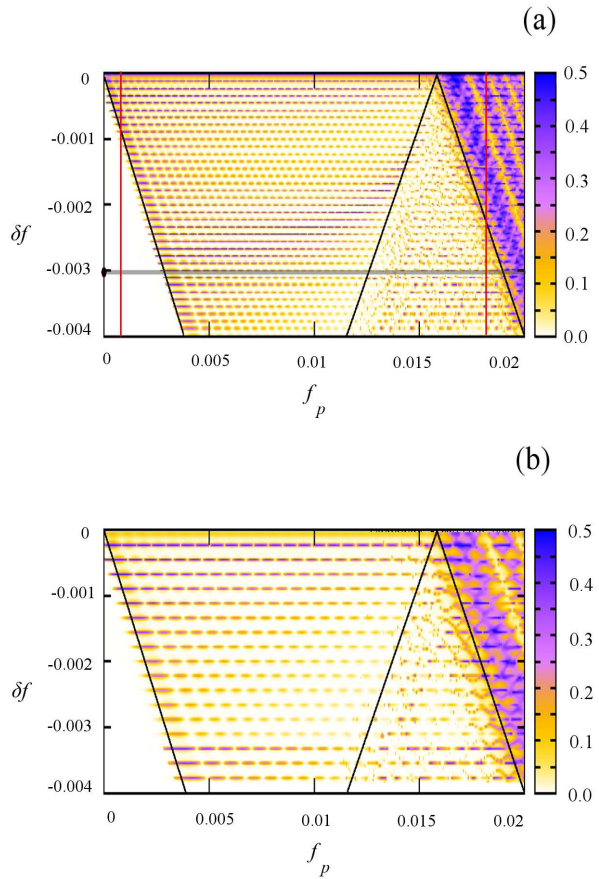}
\end{center}
\caption{(color online) Large amplitude spectroscopic  (half)
diamonds obtained for the Josephson flux qubit with $\eta=0.25$
and $\alpha=0.8$. The intensity of $\overline {P}_{-}$ is plotted
as a function of flux detuning $\delta f$ and rf amplitude $f_p$.
(a) The system is driven at a frequency $\omega=0.001$.
Calculations were done using $N_l=6$ (six levels), $150\le K\le
300$ and $M=1024$. The black solid lines indicate the edges of the
first diamond D1 and a the beginning of the second one D2. The
black dot indicates a particular value of flux detuning
$\delta f\simeq -0.003$ for $f_{p}=0$. See text for a detailed
analysis.
(b)  The flux qubit is driven at a frequency $\omega=0.002$.
Calculations were done using  $N_l=6$,  $150\le K\le 250$ and
$M=1024$. Data points correspond to a fine grid of
 $\Delta f_0=2 \times 10^{-5}$ and $\Delta f_p=3 \times 10^{-5}$
} \label{f5}
\end{figure}

Besides the time dependence,  the average occupation probability Eq.(\ref{p+a}) is the key
quantity for  the spectroscopic analysis performed in  recent experiments.  \cite{valenzuela,oliver}
 We  analyze  the
patterns of $\overline {P}_{-}$ in the $\left( f_{p}, \delta f \right)$ space.
 As we shall show below, as the amplitude  $f_{p}$  is increased, $\overline {P}_{-}$
will  exhibit  a reacher  and more involved structure.

Regarding  the  parameters employed in the numerical calculations,
for small  driving amplitude $f_p$ and  flux detuning $f_0\simeq
0.5$, the discretization grid needed  to compute   the eigenstates
$|n;f_0\rangle$ can be constructed with  quite small values of $M
\sim 128$. \cite{ferron} However, in the strong driving regime, we
need to use $M=256-1024$. In addition, we employ $K \sim 150-400$
to attain convergence in the values of   $\overline {P}_{-}$. As
we already mentioned $N_l$  is mainly determined by $f_{p}$. All
the calculations, otherwise specified, have been performed with
the lowest six levels.

In Fig. \ref{f5}(a)  we show the contour plot of  $\overline{P}_-$  as a
function of $\left( f_{p}, \delta f \right)$ for $\delta f < 0 $
and  $\omega=0.001$. The range of values of $ f_{p}$ and $\delta f$  has been
selected to explore the  region  of the energy spectrum of Fig.\ref{fig2}
containing   the avoided crossings $\Delta_{00}$, $\Delta_{10}$,
$\Delta_{11}$ and $\Delta_{20}$. The plot is obtained by calculating points
with a grid of $\Delta f_0=2 \times 10^{-5}$ and $\Delta f_p=3 \times 10^{-5}$.
A clear  pattern of maxima and minima  forming a half diamond-like structure can be observed
in Fig. \ref{f5}(a).
For the sake of clarity we have drawn  lines indicating the boundaries of the first (half) diamond  D1 and the
beginning of the second one D2.
For   $\delta f > 0 $  the diamond pattern is completed with $1- \overline{P}_-$.
The qualitative agreement  with the  diamond structure observed in Ref. \onlinecite{valenzuela} is
evident.  However,  in this coherent regime
the spectroscopic diamonds have much less 
contrast than in the experiments of Ref.\onlinecite{valenzuela}.

 In Fig. \ref{f5}(b)  we show a similar  contour plot of  $\overline{P}_-$
for a higher frequency $\omega=0.002$.
A very similar  pattern of maxima and minima  forming a half diamond-like structure can be observed
in Fig. (\ref{f5})(b). For this larger driving frequency  the distance between
 resonances is increased, as it is expected from Eq.(\ref{p2ls}).
It is important to remark that when we increase the driving frequency and thus the sweep rate, we lose
resolution in the obtained spectroscopic diamonds as it
can be checked by inspection of Fig.\ref{f5}(a) and Fig.\ref{f5}(b).

\begin{figure}[th]
\begin{center}
\includegraphics[width=20pc]{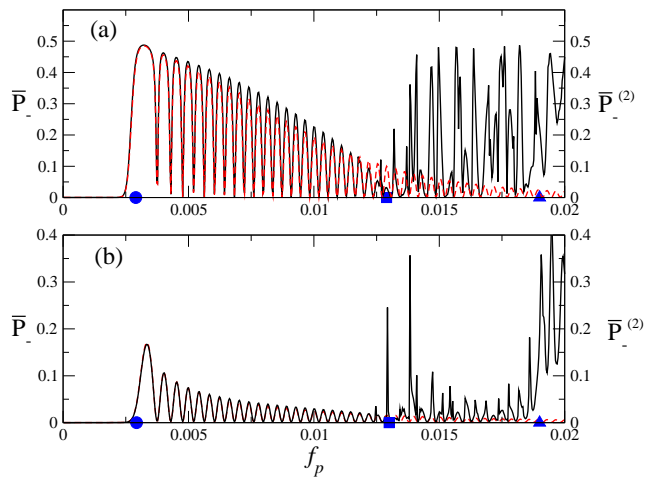}
\end{center}
\caption{(color online) $\overline{P}_{-}$ (black solid line)
calculated employing the lowest 6 levels and
$\overline{P}_{-}^{(2)}$ (red dashed line) calculated  employing the lowest
2 levels
as a function of the driving amplitude. Calculations were done
using $K=150-300$ and $M=1024$. (a) $\delta f=-0.00309$ and (b)
$\delta f=-0.00308$. The circle, square and triangle denote the
beginning of the first diamond D1, the end of D1, and the
beginning of the second diamond D2 respectively, for the present
value of $\delta f$.  See text for details.} \label{f6}
\end{figure}

Emulating the experimental protocol, in the following  we analyze
the structure of the diamonds in order to extract spectroscopic information,
focusing on a fixed frequency, $\omega=0.001$.
We start by considering a particular static flux detuning $f_0\simeq0.49691$
{\it i.e.} $\delta f\simeq -0.00309$ (black dot in
Fig.\ref{f5}(a) and vertical dashed-dotted line in Fig.\ref{fig2})
that  satifies for $\omega= 0.001$
the fast driving TLS  $n=28$ resonance condition, $\epsilon_{0}=28 \omega$.
The initial  ground state is  $|0; 0.49691\rangle$ and thus for
$f_p=0$ is  $\overline{P}_-=0$.
 As the driving amplitude is increased, the net transfer
of population over many driving periods  will
translate in a finite value of   $\overline{P}_-$.
Along the horizontal line  defined at  $\delta f\simeq -0.00309$ in
Fig. \ref{f5}(a),
the  first diamond  D1 starts at  $f_{p}^{D}=0.003$ (${D}$ labels
the parameters extracted from the diamonds). From Fig.\ref{fig2} one can
check that this value is roughly the threshold amplitude needed
to reach the  first avoided crossing $\Delta_{00}$  for the considered value
of  $f_0=0.49691$.
 For $f_p>0.003$  the multiple passages through the avoided crossing
$\Delta_{00}$ are  reflected  in the observed interference pattern
that, up to  $f_p \sim 0.01$, it is rather  well described by the
quasiperiodic behavior of the Bessel function $J_{n=28}
(4\pi\alpha E_J S_{01}f_p/ \omega)$ entering in the definition of
$\Gamma_{n=28}$ in Eq.(\ref{l2ls}). If  the driving amplitude
$f_p$ is further increased the interference patterns persist, but
the positions of the maxima and minima do not follow the Bessel
function dependence, as the description of the resonances  in
terms of the fast driving TLS formula is not accurate for these amplitudes. A
detailed evidence of this behavior is shown in Fig.\ref{f6}(a)
where we plot   a cut of  $\overline{P}_{-}$ along the horizontal
line $\delta f= -0.00309$  depicted in Fig.\ref{f5}(a),
corresponding to the condition $\varepsilon_0=28\omega$ ($n=28$ resonance
condition for the TLS in the fast driving regime), together with the same
quantity computed keeping only the lowest 2 levels,
$\overline{P}_{-}^{(2)}$, {\it i.e.} with $N_l=2$. In
Fig.\ref{f6}(b) we plot a case slightly different (slightly-off $n=28$
resonance condition for the TLS in the fast driving regime), for
$\delta f= -0.00308$, where $\overline{P}_{-}$ is clearly smaller.
For  small amplitudes, $\overline{P}_{-}$ is nicely followed  by
$\overline{P}_{-}^{(2)}$,  that indeed reproduces quite accurately
the interference patterns both on-resonance and off-resonance, in
Fig.\ref{f6}(a) and (b), respectively.
 However,  for $f_p \gtrsim 0.01$ the departure of
$\overline{P}_{-}^{(2)}$ from the actual behavior
is notorious, even before the end of the first diamond D1.
In addition,  $\overline{P}_{-}^{(2)} \rightarrow 0$  reflecting
the fact that higher levels besides the lowest two are populated as $f_p$ increases.

Unlike its beginning, the end of D1  is not so sharply defined, showing a rather
poor  change in contrast.
In the experiments of Ref.\onlinecite{valenzuela} the data show a larger reduction in contrast,
due to  fast relaxation through intra-well transitions.
Furthermore, in  the experiment  is  $\Delta_{10} \gg \Delta_{00}$
and  the population transfer is dominated by  the transition at
the avoided crossing   $\Delta_{10}$, with no explicit signatures
of additional  multiple passages  through the  extra avoided
crossing $\Delta_{11}$ (see Fig. 1(c) in
Ref.\onlinecite{valenzuela}). In our case we have verified that,
although  $\Delta_{10} = 2 \times 10 ^{-3} \gg \Delta_{00} = 3 \times
10 ^{-4} $ is roughly the same relation  than in the
Ref.\onlinecite{valenzuela}, additional  transitions at
$\Delta_{11} = 1 \times 10 ^{-2}$
 contribute to  sustain the values of  $\overline{P}_-$,
 resulting in an effective  reduction of  the  contrast
 at the end of D1.
 It is plausible  that in the experimental qubit \cite{valenzuela}, a high value of $\Delta_{11}$ gives,
 unlike our case, a negligible  transition probability at this avoided crossing.

Here, we find that despite the poor contrast in a highly coherent
DJFQ, the end of the first diamond is quite identifiable at
$f_{1e}^{D} \sim 0.013$, giving a value of $f_{10}^{D}=0.484$ in
good agreement with the position of   the second avoided crossing
$\Delta_{10}$ obtained from the analysis of the spectrum depicted
in Fig.\ref{fig2}. Thus, the boundaries of the first diamond give
a rather satisfactory  determination of the position of
$\Delta_{00}$ and   $\Delta_{10}$, respectively.

For values of $f_p> 0.013$ the competition between the different transitions at  the  avoided
crossings $\Delta_{00}$, $\Delta_{10}$ and $\Delta_{11}$,
turns the interpretation of the pattern followed by  $\overline{P}_-$ rather
complicated. However the beginning of
the second diamond  D2 at $f_{2s}^{D}=0.019$, gives the position of the  avoided crossing $\Delta_{01}$,
at $f_{01}^{D}=0.516$ very close to the exact value (see Fig.\ref{fig2}).

In analogy with the experimental analysis \cite{valenzuela}, the diamonds profiles can be studied in more
detail for a given amplitude $f_{p}$ and sweeping the flux detuning $f_0$. To this end, we
select   two vertical lines in Fig.\ref{f5} that correspond to $f_p=0.001$ and  $f_p=0.018$,
respectively.

\begin{figure}[th]
\begin{center}
\includegraphics[width=20pc]{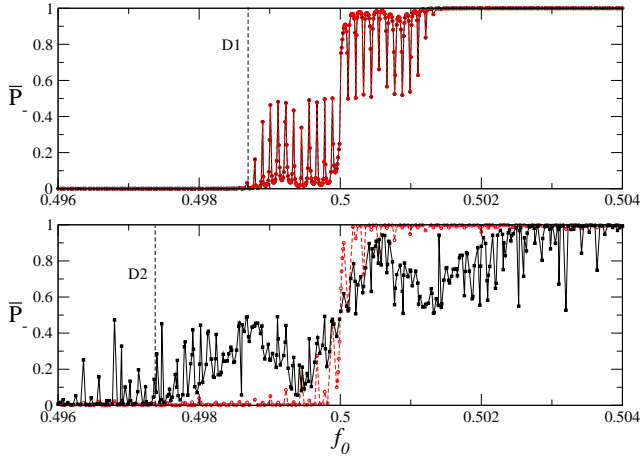}
\end{center}
\caption{(color online) $\overline{P}_{-}$ (black solid line) as a
function of the static flux  $f_0$ for two different values of the
driving amplitude $f_p=0.001$ (upper panel) and $f_p=0.018$ (lower
lanel) calculated using $N_l=6$. In both panels, a dashed line
denotes the value of $f_0$ that for $f_p=0.001$ ($f_p=0.018$)
gives  the beginning  of the first(second) diamond D1 (D2). As a
comparison we plotted in red dashed line the
$\overline{P}_{-}^{(2)}$ computed employing the lowest 2 levels.
Numerical calculations were done with $K=150-300$ and $M=1024$.
See Fig. \ref{f5}  and text for further details. } \label{f7}
\end{figure}

In  the upper panel of Fig.\ref{f7}  we show  $\overline{P}_-$  along  $f_p=0.001$.
A dashed line indicates the value  $f_0^{D1} =0.4988$ at which the vertical line defined by $f_p=0.001$
intersects the  lower edge of the D1 (see Fig.\ref{f5}). As expected, the
symmetry $\overline{P}_{-}  \rightarrow  {1- \overline{P}_{-}} $  around   $f_0=0.5$ holds.
The profile of $\overline{P}_{-}$,   sweeping
the  different  equally spaced n-resonances  as $f_{0}$ changes, seems to be in very good  agreement with the
predicted TLS  resonance pattern. Indeed we have included the results for  $\overline{P}_{-}^{(2)}$
computed employing the lowest 2 levels, which are essentially superimpossed to the actual   $\overline{P}_-$.
An estimate of the observed number of resonances  obtained employing  Eq.(\ref{l2ls})
is  $n= [\epsilon_{0}/\omega]=[4 \pi\alpha E_J  S_{01} |\delta f|/ \omega]$,
being $[...]$ the integer part.
In this case is  $|\delta f|= |0.5 -f_{0}^{D1}|= 0.00122$ and $4 \pi\alpha E_J
  S_{01} \sim 9$. Thus for $\omega=0.001$
we obtain $n=11$, which is exactly the  number of maxima displayed in the upper panel of Fig.\ref{f7} for the
selected range of flux detunings. Thus this analysis, complemented with the previous one performed in
Fig.\ref{f6}, confirms that  close to the beginning  of  the first diamond D1, the TLS description
is quite accurate.

The lower panel of Fig.\ref{f7} displays $\overline{P}_-$ for $f_p=0.018$ and  the beginning
of the second diamond D2 is indicated at  $f_{0}^{D2}=0.4974$ by the vertical dashed line.
The  erratic pattern  of resonances in    $\overline{P}_-$  is in correspondence with   the results
presented in  Fig.\ref{f6} for amplitudes inside the second diamond D2. As it occurred in that case,
the competition between the transitions at different avoided crossings  gives a profile of the occupation
probability that  strongly  departs from a simple interference pattern
as given by  Eq.(\ref{l2ls}) and/or for the pattern displayed by  $\overline{P}_{-}^{(2)}$.

Coming back to complete the spectroscopic analysis of Fig.\ref{f5}, the end of the second diamond should
be expected at $f_{p}\simeq 0.018$ corresponding to  the position of the $\Delta_{20}$ avoided crossing.
However, as it is easily checked from   Fig.\ref{f5}, we do not obtain the end of
the second diamond for this value of $f_{p}$. Indeed D2 starts for  a larger
value of  the amplitude.
In our case $\Delta_{20}\sim 1 \times 10 ^{-7} \ll \Delta_{00}$ and therefore
the transition probability $\propto {\Delta_{20}}^{2}/ \omega f_{p4} \rightarrow 0$.
As a consequence,  the spectroscopic diamond contains no visible information
on the avoided crossing  $\Delta_{20}$. This small gap should correspond to a crossing
of transverse modes, which are not easily probed by the driving $f(t)$ which
acts mainly along the longitudinal $\varphi_l$ direction. Similar drawback for
detecting transverse modes has been reported in the experiment.\cite{valenzuela}
A possible way to   increase the resolution of this gap is to
increase the transition probability by reducing the  driving frequency $\omega$.
However, larger  driving periods  could be concomitant with the
loss  in resolution of the  indivual  n-resonances. \cite{valenzuela}.

\begin{figure}[th]
\begin{center}
\includegraphics[width=20pc]{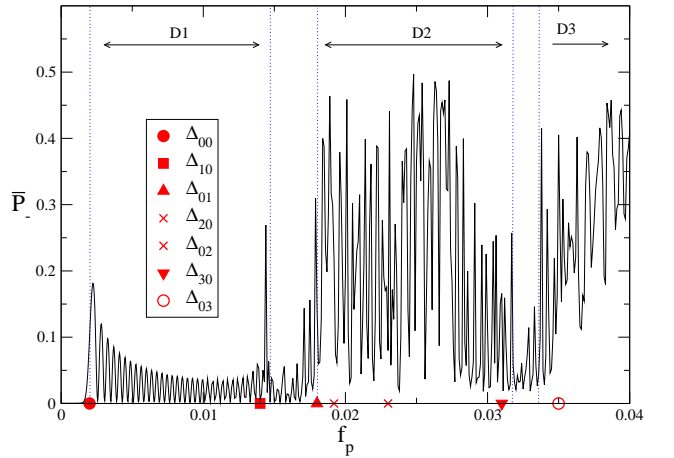}
\end{center}
\caption{(color online) $\overline{P}_-$ for $\delta f=-0.002$
 calculated employing the lowest 8 levels as a
function of the driving amplitude. Calculations were done using
$K=150-400$ and $M=1024$. The red symbols in the axis show the
values of $f_p$ for which the different avoided crossings are
reached (see Fig. 2). The vertical dotted lines indicate the borders of   
the  different spectroscopic diamonds D1, D2, D3} \label{f6d}
\end{figure}

Obtaining numerically the   diamond pattern besides the beginning
of the second one D2 is a formidable task, essentially due to the
extremely large CPU needed. Fig. \ref{f6d}  shows a cut of
$\overline{P}_-$ for $\delta f=-0.0020$  up to  amplitudes $f_p
\sim 0.04$.  For the larger amplitudes ($f_p>0.02$), we needed to
perform the calculations employing 8 levels ($N_l=8$). For the
small amplitudes the edges of the first diamond D1, and the
beginning of the second diamond D2  are clearly visible in the
abrupt changes exhibited by $\overline{P}_-$. On  the other hand,
as we have already mentioned, no evidence of population transfer
is obtained for the transverse avoided crossings $\Delta_{20}$ and
$\Delta_{02}$. For larger amplitudes, one can distinguish
a region where $\overline{P}_-$ is small ($\overline{P}_-\lesssim 0.1$) 
as separating the end of the second
diamond D2 and the beginning of the third diamond D3. 
In Fig. \ref{f6d} we show the points in $f_p$ where 
the avoided crossings at $\Delta_{30}$ (triangle down) and 
$\Delta_{03}$ (open circle) are reached.

\section{Excited State Amplitude Spectroscopy}
\label{sn2}

In this section we explore  an alternative amplitude spectroscopic method
to study the quantum dynamics of the DJFQ
starting from a different initial condition.
The {\it gedanken} experiment consist on preparing the  system in the first excited
state $|n=1,f_0\rangle$ for $f_0\lesssim 0.5$.  In this way, and depending on the value of $f_0$ chosen,
the $\Delta_{00}$  or $\Delta_{10}$  avoided crossing could be reached first as the amplitude is increased.
With this initial state, the  time averaged occupation probability is:

\begin{equation}
\overline{P}_+^e= 1 - \overline{P}_-^e =\sum_{n,m=0}^{N_l-1}p_{nm}
\Lambda^e_{nm}\; , \label{p+ae}
\end{equation}

\noindent where

\[
\Lambda^e_{nm}=\sum_{\beta=1}^{N_d}\sum_{k=-K}^{K} \langle
n,k|\phi_\beta\rangle\langle \phi_\beta|1,0\rangle \langle
1,0|\phi_\beta\rangle\langle \phi_\beta|m,k\rangle  \;.
\]

\noindent

If we compare with Eq.(\ref{p+a}) we see that now $\overline{P}_+^e$  depends on the
amplitude $ \langle \phi_\beta|1,0\rangle$ instead of $ \langle \phi_\beta|0,0\rangle $.
In general, in highly coherent devices,
one can define an average occupation probability
$\overline{P}_+^{(s)}$ depending on the initial state $|s\rangle$,
with the amplitude $ \langle \phi_\beta|s,0\rangle$ instead of $ \langle \phi_\beta|0,0\rangle $
in Eq.(\ref{p+a}).

\begin{figure}[th]
\begin{center}
\includegraphics[width=20pc]{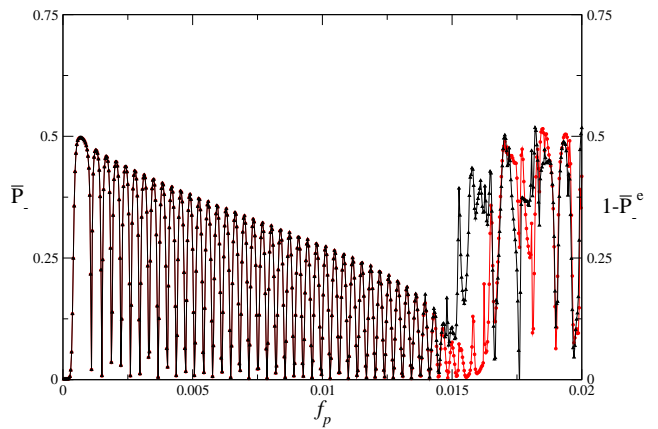}
\end{center}
\caption{(color online) $\overline{P}_{-}$ (red line) and
$1-\overline{P}_-^e$ (black line) as a function of the driving
amplitude for $\delta f=-0.00066$. Calculations were done employing
the lowest 6  levels using $K=150-300$ and $M=1024$.} \label{ex1}
\end{figure}

 For values of $f_0\lesssim 0.5$ and $f_p \rightarrow 0$ we have
$\overline{P}_+=1$ ($\overline{P}_-=0$) and $\overline{P}_+^e=0$
($\overline{P}_-^e=1$). In Fig. \ref{ex1} we plot
$\overline{P}_{-}$ and $1-\overline{P}_-^e$  as a function of the
driving amplitude for a fixed value for the detuning ($\delta
f=-0.00066 $), such that the avoided crossing $\Delta_{00}$ is
reached for smaller amplitudes than the $\Delta_{10}$. For small
values of $f_p$ the system behaves as a  TLS and  both
probabilities give the same information. As the driving amplitude
is  increased  ($f_p\sim 0.015$) noticeable differences between
$\overline{P}_-$ and $1-\overline{P}_-^e$ emerge. Indeed, in the
highly coherent case the initial condition plays an important role
in the quantum dynamics of the system already when approaching the
second avoided crossing $\Delta_{10}$ .

\begin{figure}[th]
\begin{center}
\includegraphics[width=20pc]{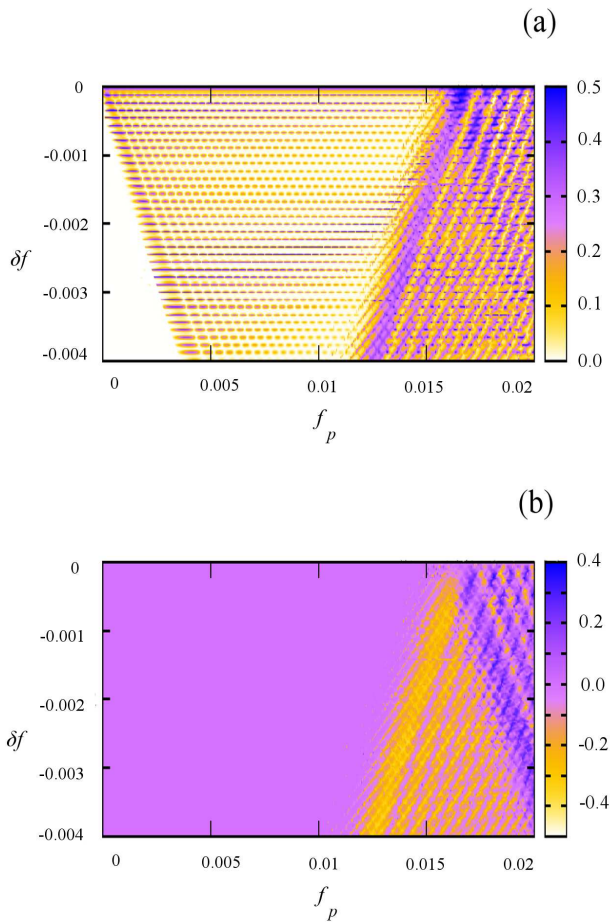}
\end{center}
\caption{(color online) Large amplitude excited state
spectroscopic (half) diamonds obtained for the  DJFQ  with
$\eta=0.25$ and $\alpha=0.8$ at a driven frequency $\omega=0.001$.
Calculations were done employing $N_l=6$ (six levels), $150\le
K\le 300$ and $M=1024$.(a) Intensity plot of
$1-\overline{P}_{-}^{e}$. (b) Intensity plot of
$\overline{P}_{-}-(1-\overline{P}_-^e)$} \label{dex1}
\end{figure}

In Fig. \ref{dex1} (a) we show the contour plot of $1-\overline{P}_{-}^{e}$
as a function of $(f_p,\delta f)$ for $\delta f<0$ and $\omega=0.001$. The
range of values of $f_p$ and $\delta f$ and the grid are the same used
to obtain the results showed in Fig. \ref{f5}.  By inspection of Fig.
\ref{f5} (a) and \ref{dex1} (a) we conclude that both amplitude
spectrospy methods give the same information before the $\Delta_{10}$
avoided crossing is reached. While the end of the first diamond $D1$ is determined
with a good  contrast with the excited state amplitude
spectroscopy method, the beginning of the second
diamond $D2$ is still very difficult to determine.

In order to analyze the information hidden in the different espectroscopic
diamonds we plot the difference between the probabilities
used to construct the diamonds in Fig. \ref{f5} (a) and Fig. \ref{dex1} (a).
In Fig. \ref{dex1} (b) we show the contour plot of
$\overline{P}_-  +\overline{P}_{-}^{e} -1$ as a function of $(f_p,\delta f)$
for $\delta f<0$ and $\omega=0.001$. As we mentioned, the difference is
zero for values of $f_p$  such that the $\Delta_{10}$ is not reached.
It is interesting to mention  that now  the end of the first diamond $D1$ and
the beginning of $D2$ can be determined with a rather good contrast, due the cancellation of some
intrincated  interference patterns  present in both $\overline{P}_-$ and $1-\overline{P}_-^ {e}$.

\begin{figure}[th]
\begin{center}
\includegraphics[width=20pc]{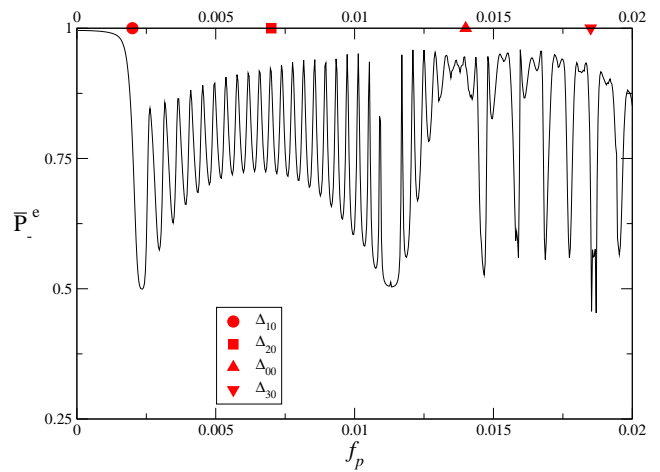}
\end{center}
\caption{(color online) $\overline{P}_{-}^e$
calculated  employing the lowest 8 levels as a function of the
driving amplitude for $\delta f=-0.014$. The system is driven at a
frequency $\omega=0.001$, $K=150-300$ and $M=1024$. The red
symbols show the position of the different avoided crossings.}
\label{ex2}
\end{figure}

Finally we evaluate the probability $\overline{P}_-^e$ as a function of
the driving amplitude for a fixed value of the flux detuning close to the
$\Delta_{10}$ avoided crossing. In this case $\overline{P}_-^e$ follows the TLS behavior
up to  the  $\Delta_{10}$ avoided crossing, but once the driving amplitude
reaches the  $\Delta_{00}$  additional levels should be included in order to properly described the
quantum dynamics of the DJFQ.
In Fig. \ref{ex2} we plot $\overline{P}_-^e$ for $f_0\sim f_{10}=0.484$, as a function of the
driving amplitude  up to values that drive the system close to the $\Delta_{30}$, for which we
have to employ eight levels in the numerical calculations.
In the figure  the  changes in  $\overline{P}_-^e$ reveal the position of the different avoided crossings
allowing a clear detection of  $\Delta_{10}$, $\Delta_{00}$ and $\Delta_{30}$.

\section{SUMMARY AND CONCLUSIONS}
\label{sc}

We have numerically solved the quantum dynamics of the device for
the JFQ under strong harmonic driving in the fully coherent
regime.  Starting from the ground state we have studied the
temporal evolution of the occupation probability and analyzed the
spectroscopic diamonds obtained for the time average occupation as
a function of flux detuning and driving amplitudes. We have shown
that for small amplitudes the description in terms of a TLS
reproduces very well the observed pattern of
Landau-Zener-St\"uckelberg interferences, as expected. On the other
hand, the TLS description breaks down for driving amplitudes such
that the avoided crossing $\Delta _{10} $ is reached. The
spectroscopic diamonds exhibit in this case interference patterns
with a rather complex structure, due to the coherent evolution
among all coupled energy levels. This situation is different from
the experiment of Ref.\onlinecite{valenzuela} where there is a
higher contrast in the diamond patterns due to the intra-well
relaxation and short coherence times. In spite of this, in the
fully coherent regime explored in this work, we find that the
edges of the diamonds clearly define the position of the different
avoided crossings, as can be observed in Fig.\ref{f6d}, for
example, even when the contrast is rather poor. In Section
\ref{sn2} we have proposed a way to obtain further information in
this case. In a highly coherent DJFQ it is possible to prepare the
system in the first excited state, for example with a $\pi$ Rabi
pulse. From there, the excited state amplitude spectroscopy could
be performed. A comparison of the occupation probabilities
obtained from the ground state amplitude spectroscopy and the
excited state amplitude spectroscopy, as performed in
Fig.\ref{dex1}(b), can now bring good contrast for the resolution
of the second diamond. In a perfectly coherent closed system one
could continue even further, performing another amplitude sweep
starting from the second excited state, compare it with the
results obtained starting from the first excited state, and so on.
Of course, in a real system the possibility of these ``excited
state amplitude spectroscopies" will be strongly limited by
decoherence and relaxation processes. In current highly coherent
DJFQ (with dephasing times of the order of 1$\mu$s) the first
excited state amplitude spectroscopy seems to be feasible. In this
case, this could give an important indication of the coherence of
the device as wells as additional and complementary information of
the multilevel structure of the energy spectrum of DJFQ.

\acknowledgments
We acknowledge discussions with Sergio Valenzuela. We also
acknowledge financial support from CNEA,  CONICET
(PIP11220080101821 and PIP11220090100051) and ANPCyT (PICT2006-483
and PICT2007-824).


\begin{thebibliography}{}
\bibitem{nota} For example in Ref.\onlinecite{noise} is $T_2 \sim 1\mu$s while
in Ref.\onlinecite{valenzuela} is $T_2\sim 20$ns.
\end{thebibliography}
\end{document}